\documentclass[12pt]{article}
\usepackage[right=1in,left=1in,top=1in,bottom=1in]{geometry}
\usepackage{amsfonts, amssymb}
\usepackage{hyperref}
\usepackage{amsmath}
\usepackage[utf8]{inputenc}
\usepackage[T1]{fontenc}
\usepackage{pxfonts}
\usepackage{graphicx}
\usepackage{setspace}

\usepackage{amsthm}

\newtheorem{theorem}{Theorem}
\newtheorem{corollary}{Corollary}

\theoremstyle{definition}

\theoremstyle{remark}

\title{\Large{Algorithmic Randomness and Physical Typicality}}

\author{Jeffrey A. Barrett\footnote{Department of Logic and Philosophy of Science, University of California, Irvine, Irvine, CA 92697-5100. Email: j.barrett@uci.edu} 
, Eddy Keming Chen\thanks{Hal{\i}c{\i}o{\u g}lu Data Science Institute and Department of Philosophy,  University of California, San Diego, 9500 Gilman Dr, La Jolla, CA 92093-0119. Email: eddykemingchen@ucsd.edu  }
~~and 
Josiah Lopez-Wild\thanks{Department of Logic and Philosophy of Science, University of California, Irvine, Irvine, CA 92697-5100. Email: jlopezwi@uci.edu}}

\date{\small{Forthcoming in Proceedings of The 2026 Meeting of the Philosophy of Science Association}}

\begin{document}

\maketitle

\begin{abstract}
Appeals to typicality are common in physics, but it is often unclear what it means for a physical state to be typical relative to a probability measure, and correspondingly unclear what a law that appeals to typicality asserts. Here we consider how one might characterize physical typicality using ideas from the theory of algorithmic randomness. As a concrete example, we show how taking a physical state to be typical relative to a computable measure when it is Martin-L\"of random allows one to formulate the distribution postulate in Bohmian mechanics as a statistical constraining law of the theory. Using a toy model, we show how this constraint guarantees the standard Born statistics for computable experimental protocols. Algorithmic Bohmian mechanics (aBM) thus illustrates how algorithmic randomness may be used to provide precise content to a statistical law.
\end{abstract}

\noindent

\vspace{3mm}

\section{Introduction}

Appeals to \textit{typicality} play an important role in physics. A theory may associate a probability measure with a space of possible worlds, histories, or initial conditions and then claim that the actual world is typical relative to that measure. But measure theory characterizes the size of sets, not the typicality of a particular world. To say that some property holds for almost all possible worlds relative to a specified measure is not to say what it is for the actual world to be typical relative to the measure. Nor can one define a typical world as one lying outside every measure-zero set. Since, in the usual continuous setting, every world lies inside some measure-zero set, no world would count as typical. 

This matters, especially, when one aims to use the notion of typicality to express the content of a physical law. If a law just says that the actual world is typical, and typicality is understood in the usual measure theoretical sense, then it is entirely unclear how the law in any way constrains the state of the actual world. In contrast, thinking of typicality in terms of algorithmic randomness allows such a law to be understood as a direct constraint on physical possibility, one that excludes a precisely defined set of worlds. Only algorithmically typical worlds are possible.

This short paper has two aims. The first is to show how the notion of typicality can be made precise using the theory of algorithmic randomness. In this, we are joining other philosophers and physicists who have sought to explicate typicality in terms of algorithmic randomness. Recent examples include Francesca Zaffora Blando (2025) in epistemology and Klaas Landsman (2023) in physics.\footnote{For the broader literature on typicality in physics, see D\"urr et al.\ (1992), Goldstein (2012), Lazarovici and Reichert (2015), Allori (2020), Hubert (2021), Wilhelm (2022), Lazarovici (2023),  and the survey in Frigg and Werndl (2024, sec.~4.5).} The intuition is that a world or state (or point in a measurable space) is \textit{typical relative to a measure} when it is not special in any effectively describable way given that measure. Consider two quick examples. The point at the center of a computable Gaussian distribution is special, and hence atypical, because it can be effectively identified given the measure. The same is true of the point having one third of the Gaussian measure to its right. In each case, the point can be effectively singled out given the measure. There are several ways to make this intuition precise.

The algorithmic notion of Martin-L\"of randomness provides a particularly natural criterion of typicality, or non-specialness, relative to a measure. The idea is to consider all uniformly effective statistical tests whose failure sets have arbitrarily small measure. A world or state (or point in a measurable space) is Martin-L\"of random relative to the measure when it passes every such test, or equivalently, when it belongs to no effectively describable measure-zero set. In this precise sense, it is typical because it is not special in an effectively describable way. 

Our second aim is to apply this notion to physical typicality and show how to use it to specify a concrete physical law. Here we go a step beyond the explication project just mentioned. Rather than using algorithmic randomness to analyze typicality claims or to sharpen almost-sure theorems, we use it to give the content of a law of nature. Specifically, we will show how to formulate the Bohmian distribution postulate as a constraining law in the context of a theory that we will call \textit{algorithmic Bohmian mechanics} (aBM).

Standard Bohmian mechanics is a deterministic theory that reproduces the empirical predictions of the standard collapse formulation of quantum mechanics. Since there are no stochastic processes in the theory, forward-looking probabilities are often regarded as epistemic---they reflect our ignorance of initial conditions rather than any objective indeterminism. As in classical statistical mechanics, the probabilities arise from a special statistical boundary condition. In Bohmian mechanics, this condition is the \textit{distribution postulate}.

The distribution postulate says that the initial state is \textit{typical relative to the standard Born measure} at an initial time $t_0$. And this is usually expressed as a stipulation that the epistemic probability density for the particle configuration $Q$ at time $t_0$ is given by $\rho(Q,t_0)=|\Psi(Q,t_0)|^2$, where $\Psi(Q,t_0)$ is the initial quantum state of the universe. But how are we to understand this stipulation? Standard presentations leave the status of the Bohmian distribution postulate unclear: is it a definition of typicality, a norm of rational credence, or an independent law of nature? The distinction matters, since each interpretation makes a different claim about the actual configuration. 

David Albert sought to explicate what it means for the initial state to be typical by suggesting that we imagine God determining the initial particle configuration using ``some genuinely \textit{probabilistic} procedure'' such that the probability of it being placed in any particular region is $|\Psi|^2$ over that region (1992, 138). But this does not really help in understanding what it means for the initial state to be typical. Since being typical in this sense is not an intrinsic property of the state, this explication \textit{depends entirely on the procedure} by which the initial state is selected (also see Landsman 2022). Inasmuch as it is unclear how we are to understand such a genuinely probabilistic procedure physically or theologically, it is unclear how we should understand the law, its associated chances, or the sense or extent to which these chances are objective.

In contrast, we will show how one might understand the distribution postulate as an \textit{algorithmic constraining law}. The law does not describe a stochastic process by which the configuration is selected. Rather, it directly restricts the physically possible initial configurations to those that are typical in that they are not exceptional in any effectively describable way relative to the quantum measure. It says, specifically, that the initial particle configuration is Martin-L\"of random with respect to the Born measure $|\Psi|^2$.\footnote{This approach builds on a literature applying algorithmic randomness to probabilistic laws. The notion of an $L^\star$ law---a law that constrains the actual world to be algorithmically random relative to a specified measure---was developed in Barrett and Chen (2025a). For a discussion of how $L^\star$ laws, together with exchangeability, yield the Principal Principle, see Barrett and Chen (2025b). The approach is also related to several recent works on algorithmic randomness in the philosophy of science. See Belot (2023, 2024), and Mierzewski and Zaffora Blando (2024) in addition to the papers mentioned earlier. For recent examples of constraint accounts of laws, see Adlam (2022), Chen and Goldstein (2022).} 

This law is part of algorithmic Bohmian mechanics (aBM), a theory we characterize using a toy model involving sequences of spin measurements. The model clarifies both the promise and the challenges of using algorithmic typicality to formulate a physical law. We then state the main result: for any uniformly layerwise computable sequence of measurements, if the actual initial configuration is Martin-L\"of random relative to the Born measure, then the resulting outcome sequence is guaranteed to be Martin-L\"of random relative to the induced push-forward measure. Throughout the paper, we assume that the initial universal wave function $\Psi$ is computable and that configuration space equipped with the Born measure is a computable probability space; by the \textit{Born statistics} we mean the probability distributions on measurement outcomes induced by the $|\Psi|^2$ measure under the relevant experimental protocols. 
For every such experimental protocol, the actual outcome sequence is therefore algorithmically typical with respect to the induced outcome measure, which gives the corresponding Born statistics. The full technical treatment, including proofs, is developed in Barrett et al. (2026).\footnote{While the two cases are not precisely analogous, a similar approach can be used to formulate the special boundary conditions required for predictive compatibility between classical statistical mechanics and thermodynamics. For recent work along these lines, see Dekkers and Landsman (2025).}

\section{Bohmian mechanics and the distribution postulate}

In its simplest form, Bohmian mechanics is characterized by four principles.\footnote{This description follows Bell (1987, 127) rather than Bohm's quantum-potential formulation. See also the survey in Goldstein (2025).} The first is a \textit{state description}: the complete physical state at a time is given by the wave function $\Psi$ and the determinate particle configuration $Q$. Second, the \textit{wave dynamics}: the wave function evolves according to the Schr\"{o}dinger equation,
\begin{equation}
i \hbar \frac{\partial \Psi}{\partial t} = \hat{H} \Psi.
\end{equation}
Third, the \textit{particle dynamics}: particles move according to
\begin{equation}
\frac{dQ_k}{dt} = \frac{\hbar}{m_k} \frac{\text{Im}(\Psi^* \nabla_k \Psi)}{\Psi^* \Psi}\Big|_Q
\end{equation}
where $m_k$ is the mass of particle $k$. One can picture the point representing the $N$-particle configuration in $3N$-dimensional configuration space as being carried along by the probability currents generated by the evolving wave function. Fourth, the \textit{distribution postulate}: at an initial time $t_0$, $\rho(Q, t_0) = |\Psi(Q, t_0)|^2$.  This is the assumption that the initial state is typical (D\"urr et al. 1992). 

A crucial dynamical fact is that the probability density $\rho = |\Psi|^2$ satisfies the continuity equation
\begin{equation}
\frac{\partial \rho}{\partial t} + \text{div}(\rho \, v^\Psi) = 0.
\end{equation}
This means that if $\rho(t_0) = |\Psi(t_0)|^2$, then $\rho(t) = |\Psi(t)|^2$ at all later times---the Born distribution is \textit{equivariant} under the dynamics. The distribution postulate thus needs to be satisfied only once, at the initial time.

In Bohmian mechanics, besides the universal configuration $Q$, only the universal wave function $\Psi$ is fundamental. A subsystem has a wave function derivatively, as its conditional wave function (D\"urr et al. 1992). To begin, we will consider an unentangled product state where each particle has a well-defined wave packet $\psi$ and the Born measure factorizes accordingly.

A simple example illustrates how the postulate works. Consider a spin-$1/2$ particle $e$ in initial state
$$
|\!\uparrow_z\rangle_e |\psi\rangle_e = \frac{1}{\sqrt{2}}(|\!\uparrow_x\rangle_e + |\!\downarrow_x\rangle_e)|\psi\rangle_e
$$
where $|\psi\rangle_e$ is a spherically symmetric wave packet of uniform probability density. When the wave packets $|\!\uparrow_x\rangle_e |\psi_{\text{up}}\rangle_e$ and $|\!\downarrow_x\rangle_e |\psi_{\text{down}}\rangle_e$ are deflected by the Stern-Gerlach magnets, the particle's trajectory depends on its initial position: if it starts in the top half of its wave packet, probability currents carry it upward into the spin-up branch; if it starts in the bottom half, it is carried downward. The measurement result is fully determined by the initial position. The distribution postulate ensures that each outcome has probability $1/2$.

\begin{figure}[!ht]
  \centering
    \includegraphics[scale=.6]{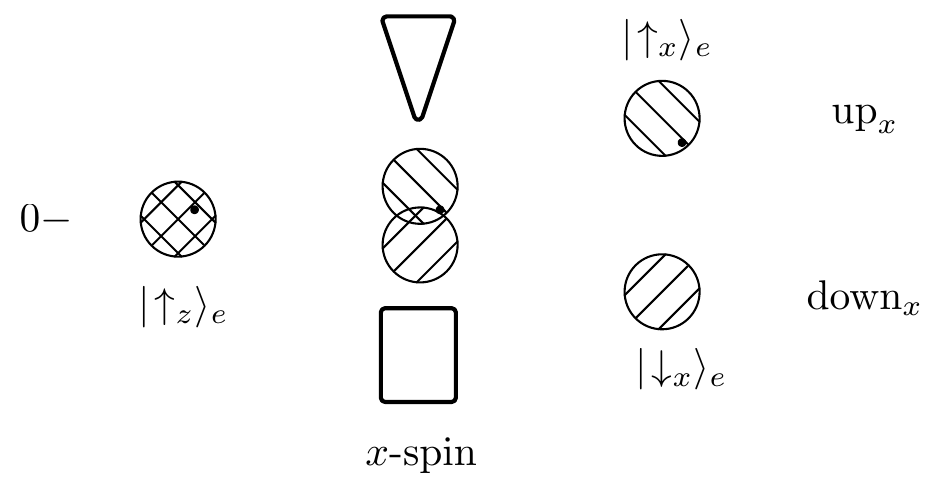}
  \caption{Stern-Gerlach measurement of $x$-spin. Whether the particle is deflected up or down depends on whether it starts in the top or bottom half of its wave packet.}
\end{figure}

The role of the distribution postulate is to ensure that probabilities over initial configurations induce probabilities for measurement outcomes that agree with the Born rule. But saying that $\rho = |\Psi|^2$ leaves open the question of what kind of constraint this is. The theory of algorithmic randomness allows us to replace this opaque statistical posit with a sharp, objective condition.

\section{Algorithmic randomness}

The key notion we need is Martin-L\"{o}f randomness (Martin-L\"{o}f 1966).\footnote{See Dasgupta (2011) and Li and Vit\'{a}nyi (2019) for introductions to algorithmic randomness.} A \textit{Martin-L\"{o}f test} is a sequence $\{U_n\}_{n \in \omega}$ of uniformly $\Sigma^0_1$ classes (constructively specifiable sets) such that $\mu(U_n) \leq 2^{-n}$, where $\mu$ is the Lebesgue measure. Each test corresponds to a way that a sequence might be ``special''---that is, might exhibit a computable pattern. A binary $\omega$-sequence $\sigma$ is \textit{Martin-L\"{o}f random} if it passes every such test: $\sigma \notin \bigcap_n U_n$ for every Martin-L\"{o}f test $\{U_n\}$.

Martin-L\"{o}f randomness captures several intuitive desiderata simultaneously. If $\sigma$ is (unbiased) Martin-L\"{o}f random, then (i) the limiting relative frequencies of $0$'s and $1$'s are each $1/2$, (ii) $\sigma$ passes every effective test for statistical independence, (iii) all finite initial segments are $c$-incompressible by a prefix-free Turing machine (for some constant $c$), and (iv) no computable betting strategy (constructive martingale) succeeds on $\sigma$. Moreover, almost all sequences are (unbiased) Martin-L\"{o}f random in Lebesgue measure. Martin-L\"{o}f randomness thus unifies standard intuitions concerning relative frequency, independence, incompressibility, and typicality into a single rigorous concept.

The notion generalizes naturally from sequences to points relative to a specified measure in an arbitrary computable probability space. A point $x$ in a computable Polish space $X$ equipped with a computable probability measure $\mu$ is \textit{$\mu$-Martin-L\"{o}f random} if it lies outside every effectively describable $\mu$-null set. We write $\mathsf{MLR}^{\mu}$ for the set of $\mu$-Martin-L\"{o}f random points. This generalization is what will allow us to state the distribution postulate for arbitrary computable wave functions and configuration spaces.

\section{A toy model}

We now develop the central idea through two examples of increasing sophistication. Consider an infinite sequence of spin-$1/2$ particles $S_1, S_2, \ldots$ each in an eigenstate of $z$-spin with a spherically symmetric wave packet $|\psi\rangle$ of uniform probability density:
$$
|\!\uparrow_z\rangle_{S_k} |\psi\rangle_{S_k} = \frac{1}{\sqrt{2}}(|\!\uparrow_x\rangle_{S_k} + |\!\downarrow_x\rangle_{S_k})|\psi\rangle_{S_k}.
$$
Suppose we plan to measure the $x$-spin of each particle in turn, recording $1$ for spin-up and $0$ for spin-down.\footnote{The toy model uses infinitely many particles, so that the outcomes form an infinite binary sequence. The general theory in \S6 does not need this idealization, since the configuration of a universe of $N$ particles is a single point of $\mathbb{R}^{3N}$, to which Martin-L\"{o}f randomness applies directly. Only the measurement sequence remains infinite.}

\subsection*{Example 1: A simple $x$-spin sequence}

Consider a law $L^\star$ that stipulates the initial particle positions using a Martin-L\"{o}f random binary sequence $\sigma_{\text{MLR}} = (1, 0, 1, \ldots)$: if the $k$th digit is $1$, particle $S_k$ starts in the top half of its wave packet; if $0$, in the bottom half.

\begin{figure}[!ht]
  \centering
    \includegraphics[scale=.6]{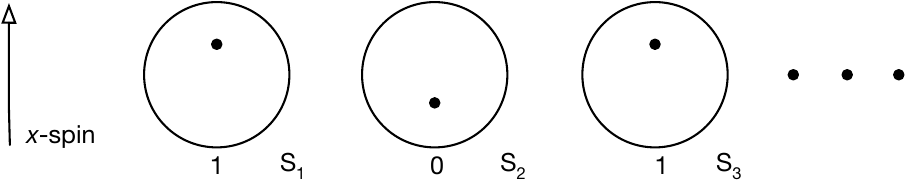}
  \caption{Particles placed in their wave packets according to the terms of a Martin-L\"{o}f random sequence.}
\end{figure}

Since whether a particle is deflected up or down by the Stern-Gerlach apparatus depends on whether it starts in the top or bottom half, the sequence of $x$-spin measurement results simply recapitulates $\sigma_{\text{MLR}}$.\footnote{Landsman mentions a suggestion by Jos Uffink that is structurally equivalent to this example (2023, section 6).} Because $\sigma_{\text{MLR}}$ is Martin-L\"{o}f random, the outcome sequence has limiting relative frequency $1/2$ for each result---and this holds with \textit{certainty}, not merely with probability~$1$. The distinction matters: probability-$1$ claims are consistent with the actual world being in the measure-zero exception set, while Martin-L\"{o}f randomness rules out every effectively describable exception.

\subsection*{Example 2: Extending to arbitrary spin directions}

Example~1 only guarantees the right statistics for $x$-spin measurements. For other spin observables, the same initial configuration may yield nonrandom results. We need a placement scheme that works for all spin directions in the $xy$-plane.

Consider another $\sigma'_{\text{MLR}} = (1, 0, 0, 1, 0, 1, \ldots)$, now interpreted recursively. The \textit{first} digit determines whether $S_1$ is in the top or bottom half of its wave packet ($1 \Rightarrow$ top). The next \textit{two} digits refine: $S_1$ is shifted left or right within its half ($0 \Rightarrow$ left), and $S_2$ is placed top or bottom ($0 \Rightarrow$ bottom). The next \textit{three} digits refine further: $S_1$ is shifted right or left within its current quarter ($1 \Rightarrow$ right), $S_2$ is shifted left or right ($0 \Rightarrow$ left), and $S_3$ is placed top or bottom ($1 \Rightarrow$ top). And so on.

\begin{figure}[!ht]
  \centering
    \includegraphics[scale=.6]{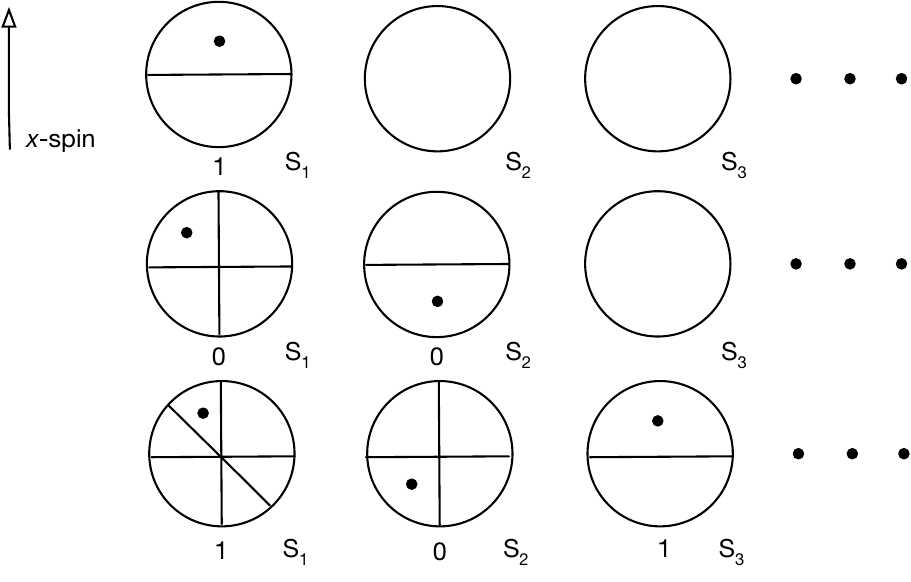}
  \caption{Recursive placement of particles using terms of a Martin-L\"{o}f random sequence. In the limit, particles are uniformly distributed over all angles.}
\end{figure}

In the limit, each particle is placed uniformly with respect to the angle of spin measurement. As a result, any computably specified spin observable in the $xy$-plane yields a Martin-L\"{o}f random sequence of outcomes with the standard quantum limiting frequency of $1/2$ for each result.

\section{Three problems and generalizations}

The toy model reveals three challenges that must be addressed to obtain a fully general theory.

\subsection*{Problem 1: Sneaky measurement sequences}

Even if the particles are distributed randomly over their wave packets, an adversary (playing a role analogous to Maxwell's demon) who knew the exact positions could choose a sequence of spin measurement \textit{directions} to guarantee spin-up every time---simply orient each Stern-Gerlach device so that the particle's offset from center lies in the ``up'' region. The resulting outcome sequence would be maximally nonrandom.

This means no placement scheme can guarantee random outcomes for \textit{all possible} sequences of measurements. The solution is to restrict attention to \textit{computable} sequences of measurements. The idea is natural: any experiment that can be finitely specified, implemented, and repeated corresponds to a computable procedure. An adversary exploiting knowledge of exact particle positions to choose measurement directions would need access to non-computable information about the initial configuration (since the positions are Martin-L\"{o}f random and hence algorithmically incompressible). The theory guarantees Born statistics for every experiment that could actually be carried out, while allowing that non-computable ``experiments'' might yield nonrandom results.

Is there a precise way to understand this computability constraint? Recent work on algorithmic randomness bears on this question (see surveys by Hoyrup (2020) and Hoyrup and Rute (2021)). The right technical notion turns out to be \textit{layerwise computability}, an effective analogue of measurability developed by Hoyrup and Rojas (2009). A $\mu$-layerwise computable function is one that is computable on each ``level'' of the Martin-L\"{o}f random points, where levels are determined by randomness deficiency. Crucially, layerwise computable functions preserve Martin-L\"{o}f randomness: if $f$ is $\mu$-layerwise computable and $Q$ is $\mu$-Martin-L\"{o}f random, then $f(Q)$ is Martin-L\"{o}f random with respect to the push-forward measure (Hoyrup and Rojas 2009; see our Theorem~1 in \S6). This is the key technical fact underlying the general theory.

\subsection*{Problem 2: Beyond the toy model}

Our examples involve idealized uniform, symmetric, separable wave packets and a single type of observable (spin). A general theory must handle arbitrary computable wave functions (including non-uniform and entangled states) and arbitrary observables (not just spin). The solution is to formulate the distribution postulate directly in terms of points in configuration space: instead of encoding positions via digits of a random sequence, one requires that the actual initial configuration $Q_0 \in \mathbb{R}^{3N}$ is Martin-L\"{o}f random with respect to the Born measure $\mu_\Psi$ on configuration space. The mathematical framework for Martin-L\"{o}f randomness on computable Polish spaces, developed in the computable measure theory literature (Hoyrup and Rute 2021), provides exactly what is needed.

\subsection*{Problem 3: From limiting frequencies to single-case credences}

In the iid setting considered here, Martin-L\"{o}f randomness guarantees a random sequence with the right limiting relative frequencies for layerwise computable measurements, but limiting statistics do not by themselves entail single-shot probabilities. One usually moves from physical chance to credence via something like Lewis's Principal Principle (1980). In the present context, a more direct route is available: if an agent's credences over measurement outcomes are \textit{exchangeable}---she does not believe the order of measurements affects the probabilities---then de Finetti's representation theorem, together with the limiting frequencies guaranteed by Martin-L\"{o}f randomness, commits her to the standard single-case Born credences. The exchangeability property of the credences encodes the agent's inductive commitment about the uniformity of nature. (See Barrett and Chen (2025b) for a more detailed discussion of the connection between $L^\star$ laws, exchangeability, and the Principal Principle.) The theory supports exchangeability in the present setup, since the systems are identically prepared and non-interacting. So the move from frequencies to credences is secured by a modest and well-motivated assumption about induction (exchangeability).

\section{Algorithmic Bohmian Mechanics}

We are now in a position to state the general theory. Algorithmic Bohmian mechanics (aBM) retains the state description, the wave dynamics, and the particle dynamics of standard Bohmian mechanics; the distribution postulate is replaced. Let $X = \mathbb{R}^{3N}$ be the configuration space of the total system and let
$$
\mu_\Psi(A) = \int_A |\Psi(Q)|^2 \, d\lambda(Q)
$$
be the Born measure induced by a computable wave function $\Psi$. Assuming $(X, \mu_\Psi)$ is a computable probability space, the distribution postulate becomes:

\begin{quote}
\textbf{Algorithmic Distribution Postulate:} The actual initial configuration $Q_0 \in X$ is Martin-L\"{o}f random with respect to the Born measure $\mu_\Psi$. That is, $Q_0 \in \mathsf{MLR}^{\mu_\Psi}$.
\end{quote}

This says that the actual initial configuration lies outside every effectively describable $\mu_\Psi$-null set. It is not exceptional in any effectively describable way relative to the measure. To borrow Albert's metaphor: God chooses a configuration that is typical relative to the standard quantum measure $\mu_\Psi$ in every computable sense.

The main result of the general theory is the following.\footnote{Full proofs are in Barrett et al. (2026). The argument relies on the Hoyrup-Rojas preservation theorem for layerwise computable maps and Kolmogorov's extension theorem.}

\begin{theorem}
Let $(X, \mu_\Psi)$ be the configuration space equipped with the Born measure. Let $\{f_n\}$ be a uniformly layerwise computable sequence of functions, and let $\mu_\omega$ be the induced joint measure on measurement outcomes, which gives the Born statistics for this measurement sequence. If $Q_0 \in X$ is $\mu_\Psi$-Martin-L\"{o}f random, then the outcome sequence $(f_1(Q_0), f_2(Q_0), \ldots)$ is $\mu_\omega$-Martin-L\"{o}f random.
\end{theorem}

In other words, a Martin-L\"{o}f random initial configuration, when subjected to any uniformly layerwise computable sequence of measurements, yields a Martin-L\"{o}f random sequence of outcomes with respect to the induced joint outcome measure $\mu_\omega$. As an immediate consequence in the iid case:

\begin{corollary}
Suppose that $\{f_n\}$ is a sequence of uniformly layerwise-computable functions representing repetitions of the same binary-outcome measurement on independently and identically prepared, non-interacting systems, so that the induced outcome measure $\mu_\omega$ is a computable iid measure.  If $Q_0$ is $\mu_\Psi$-Martin-L\"{o}f random, then
$$
\lim_{n \to \infty} \frac{1}{n} \sum_{i=1}^n f_i(Q_0) = \mathbb{E}_{\mu_\Psi}[f_1].
$$
\end{corollary}

For these binary iid measurements, the outcome frequencies converge to the Born expectations with certainty. Moreover, this holds for every infinite computable subsequence of measurements and every observable measured infinitely often under the same iid conditions. More generally, the theory guarantees Martin-L\"of random outcome sequences for any finitely specifiable, executable, repeated experiment, with convergence to the corresponding Born expectations whenever an appropriate effective strong law of large numbers applies. 

A layerwise computable measurement in Bohmian mechanics has a natural physical interpretation. Consider an $x$-spin measurement. The Bohmian dynamics generates a deterministic flow $\Phi^{\Psi_0}_{t_1}: X \to X$ from the initial configuration to the post-measurement configuration. A partition map $g: X \to \{0, 1\}$ reads off the pointer position (spin-up or spin-down). The single measurement is the composition $f(Q_0) = g(\Phi^{\Psi_0}_{t_1}(Q_0))$, which is layerwise computable when the wave function and Hamiltonian are computable. Equivariance ensures that the Bohmian flow pushes the initial Born measure forward to the Born measure at the time of measurement; the outcome measure induced by $f$ is the further push-forward of that measure under $g$. A computable sequence of such measurements can be combined into a single layerwise computable function $f_\omega: X \to \mathbb{R}^\omega$, and the theorem guarantees that randomness is preserved.

\section{Discussion}

First, standard treatments of the distribution postulate show that for $|\Psi|^2$-almost all initial configurations, the Born statistics obtain. But ``almost all'' leaves open whether the \textit{actual} configuration might be in the measure-zero exception set. Algorithmic Bohmian mechanics strengthens this: the Born statistics hold with \textit{certainty} for every Martin-L\"{o}f random initial configuration under a uniformly layerwise computable sequence of measurements, and the algorithmic distribution postulate ensures that the actual configuration is Martin-L\"{o}f random. The exception set---the non-random configurations---is not merely measure-zero but effectively null, as an algorithm can generate sets containing all of them with arbitrarily small Born measure. The algorithmic distribution postulate rules out these configurations as physically impossible. This gives the theory a stronger empirical guarantee than standard typicality arguments provide.

Second, the algorithmic distribution postulate can be understood as an objective constraining law (which can be understood, for example, in the minimal primitivism account of laws developed by Chen and Goldstein 2022): it says that the actual initial configuration of the universe is constrained to be a member of $\mathsf{MLR}^{\mu_\Psi}$, which is a perfectly well-defined set. The constraint does not invoke a stochastic process at all. Probabilities enter when an agent with exchangeable credences uses de Finetti's theorem to extract single-case Born credences from the limiting frequencies guaranteed in the iid case considered here (Barrett and Chen 2025b). This provides a clean separation between the objective content of the law (algorithmically constrained initial conditions plus deterministic dynamics) and the epistemic role of probability (rational credences).

Third, the restriction to computable measurements is not merely a technical convenience. It reflects a substantive hypothesis about the scope of physical experimentation: experimental protocols that can be described, implemented, and repeated are computable procedures. The ``sneaky'' measurement sequences that defeat the distribution postulate are non-computable processes that require oracular access to the exact initial configuration. Algorithmic Bohmian mechanics thus draws a principled line between the experiments for which it guarantees Born statistics (all computable ones) and those for which it does not (non-computable ones that can never be effectively specified).  

Finally, in standard Bohmian mechanics, the distribution postulate floats somewhat between a law, a definition of typicality, and a recommendation for credences; in algorithmic Bohmian mechanics, its status is clear. It is a constraining law that specifies the initial condition of the universe, much as the Past Hypothesis specifies the initial macrostate in statistical mechanics, but with the precision that algorithmic randomness affords. The approach generalizes to other theories beyond Bohmian mechanics: wherever a physical theory requires a statistical boundary condition, one can ask whether that condition can be formulated as an algorithmic constraint, and what is gained by doing so. 

\section{Conclusion}

Appeals to typicality are common in physics, but it is often unclear what they mean. We have argued that algorithmic randomness allows one to specify precisely what it means for a physical state to be typical. A natural criterion is to take a state to be typical relative to a computable measure when it is Martin-L\"of random with respect to that measure. Such a state is not exceptional in any effectively describable way relative to the measure. We have shown how this criterion can specify the content of a physical law. Understood in this way, the Bohmian distribution postulate is no longer a vague statistical posit but a sharp constraint on which initial configurations are physically possible. The initial-condition law of the resulting algorithmic Bohmian mechanics simply says that $Q_0 \in \mathsf{MLR}^{\mu_\Psi}$. Together with the dynamics, this law guarantees that the actual outcome sequence is Martin-L\"of random with respect to the induced outcome measure for every computable experimental procedure. Unlike other approaches to the distribution postulate, this law says something precise about the actual physical world. 

\section*{Acknowledgements}
The authors thank David Albert and Klaas Landsman for helpful discussions. Claude was used to provide editorial advice and shorten a description of the setup and formal results of a significantly longer and more technical draft written by the authors. The authors then extensively revised the shortened draft for accuracy, argument structure, and exposition. EKC is supported by Grant 63209 from the John Templeton Foundation.


\begin{center}
\large{Bibliography}
\end{center}

\vspace{.15cm}
\noindent
Adlam, Emily. (2022) ``Laws of Nature as Constraints,'' \emph{Foundations of Physics}, \textbf{52}, 28. \\ \href{https://doi.org/10.1007/s10701-022-00546-0}{https://doi.org/10.1007/s10701-022-00546-0}

\vspace{.15cm}
\noindent
Albert, David Z. (1992) \emph{Quantum Mechanics and Experience}, Cambridge, MA: Harvard University Press.

\vspace{.15cm}
\noindent
Allori, Valia. (2020) ``Some Reflections on the Statistical Postulate: Typicality, Probability, and Explanation between Deterministic and Indeterministic Theories,'' in Valia Allori (ed.), \emph{Statistical Mechanics and Scientific Explanation: Determinism, Indeterminism, and Laws of Nature}, Singapore: World Scientific, pp.~65--111.

\vspace{.15cm}
\noindent
Barrett, Jeffrey A. and Chen, Eddy Keming. (2025a) “Algorithmic Randomness and Probabilistic Laws,” \emph{The British Journal for the Philosophy of Science}, forthcoming.  \\ \href{https://arxiv.org/abs/2303.01411}{https://arxiv.org/abs/2303.01411}

\vspace{.15cm}
\noindent
Barrett, Jeffrey A. and Chen, Eddy Keming. (2025b) “Algorithmic Randomness, Exchangeability, and the Principal Principle,” \emph{The British Journal for the Philosophy of Science}, forthcoming.  \\ \href{https://arxiv.org/abs/2510.24054}{https://arxiv.org/abs/2510.24054 }

\vspace{.15cm} 	
\noindent
Barrett, Jeffrey A., Chen, Eddy Keming, and Lopez-Wild, Josiah. (2026) “The Distribution Postulate in Algorithmic Bohmian Mechanics.”   \\ \href{https://arxiv.org/abs/2606.16165}{https://arxiv.org/abs/2606.16165}

\vspace{.15cm}
\noindent
Bell, John S. (1987) \emph{Speakable and Unspeakable in Quantum Mechanics}, Cambridge: Cambridge University Press.

\vspace{.15cm}
\noindent
Belot, Gordon. (2024) “Unprincipled,” \emph{The Review of Symbolic Logic}, \textbf{17}(2), pp.~435--474. \\
\href{https://doi.org/10.1017/S1755020323000151}{https://doi.org/10.1017/S1755020323000151}

\vspace{.15cm}
\noindent
Belot, Gordon. (2023) “That Does Not Compute: David Lewis on Credence and Chance,” \emph{Philosophy of Science}, \textbf{90}(5), pp. 1130-1139 \ \href{https://doi.org/10.1017/psa.2023.11}{https://doi.org/10.1017/psa.2023.11}

\vspace{.15cm}
\noindent
Chen, Eddy Keming and Sheldon Goldstein (2022) ``Governing without a Fundamental Direction of Time: Minimal Primitivism about Laws of Nature,'' in Yemima Ben-Menahem (ed.), \emph{Rethinking the Concept of Law of Nature}, Springer, pp.21-64. 

\noindent
\href{https://arxiv.org/abs/2109.09226}{https://arxiv.org/abs/2109.09226}

\vspace{.15cm}
\noindent
Dasgupta, Abhijit (2011) ``Mathematical Foundations of Randomness,'' in Prasanta Bandyopadhyay and Malcolm Forster (eds.), \emph{Philosophy of Statistics (Handbook of the Philosophy of Science: Volume 7)}, Amsterdam: Elsevier, pp. 641–710. 

\vspace{.15cm}
\noindent
de Finetti, Bruno. (1937) “La pr\'evision: ses lois logiques, ses sources subjectives,” \emph{Annales de l’Institut Henri Poincar\'e}, \textbf{7}(1), 1–68.

\vspace{.15cm}
\noindent
de Finetti, Bruno. (1974–1975) \emph{Theory of Probability}, Vols. 1 and 2. Trans. A. Machi and A. F. M. Smith. New York: Wiley.

\vspace{.15cm}
\noindent
Dekkers, Nino and Klaas Landsman. (2025) ``Irreversibility and randomness.'' \textit{PhilSci Archive}, https://philsci-archive.pitt.edu/27619/1/Nino-Klaas-final.pdf

\vspace{.15cm}
\noindent
D\"urr, Detlef, Sheldon Goldstein, and Nino Zangh\`i. (1992) ``Quantum Equilibrium and the Origin of Absolute Uncertainty,'' \emph{Journal of Statistical Physics}, \textbf{67}, pp.~843--907.

\vspace{.15cm}
\noindent
Frigg, Roman and Charlotte Werndl. (2024) ``Philosophy of Statistical Mechanics,'' \emph{The Stanford Encyclopedia of Philosophy} (Spring 2024 Edition), Edward N. Zalta and Uri Nodelman (eds.). \\ \href{https://plato.stanford.edu/archives/spr2024/entries/statphys-statmech/}{https://plato.stanford.edu/archives/spr2024/entries/statphys-statmech/}

\vspace{.15cm}
\noindent
Goldstein, Sheldon (2025) ``Bohmian Mechanics,'' \emph{The Stanford Encyclopedia of Philosophy}, (Fall 2025 Edition), Edward N. Zalta \& Uri Nodelman (eds.), \\ \href{https://plato.stanford.edu/archives/fall2025/entries/qm-bohm/}{https://plato.stanford.edu/archives/fall2025/entries/qm-bohm/}

\vspace{.15cm}
\noindent
Goldstein, Sheldon. (2012) ``Typicality and Notions of Probability in Physics,'' in Yemima Ben-Menahem and Meir Hemmo (eds.), \emph{Probability in Physics}, Berlin: Springer, pp.~59--71.

\vspace{.15cm}
\noindent
H\'ajek, Alan (2023) "Interpretations of Probability", in Edward N. Zalta \& Uri Nodelman (eds.) \emph{The Stanford Encyclopedia of Philosophy (Winter 2023 Edition)}. 

\noindent
\href{https://plato.stanford.edu/archives/win2023/entries/probability-interpret/}{https://plato.stanford.edu/archives/win2023/entries/probability-interpret/}

\vspace{.15cm}
\noindent
Hoyrup, Mathieu. "Algorithmic randomness and layerwise computability." Algorithmic Randomness: Progress and Prospects, edited by Johanna NY Franklin and Christopher P. Porter, Lecture Notes in Logic, Cambridge University Press, Cambridge (2020): 115-133.

\vspace{.15cm}
\noindent
Hoyrup, M., and Rojas, C. (2009). Applications of effective probability theory to Martin-Löf randomness. In International colloquium on automata, languages, and programming (pp. 549-561). Berlin, Heidelberg: Springer Berlin Heidelberg.

\vspace{.15cm}
\noindent
Hoyrup, M., and Rute, J. (2021). Computable measure theory and algorithmic randomness. In Handbook of Computability and Complexity in Analysis (pp. 227-270). Cham: Springer International Publishing.

\vspace{.15cm}
\noindent
Hubert, Mario. (2021) ``Reviving Frequentism,'' \emph{Synthese}, \textbf{199}, pp.5255--5284. \\ \href{https://doi.org/10.1007/s11229-021-03024-8}{https://doi.org/10.1007/s11229-021-03024-8}

\vspace{.15cm}
\noindent
Landsman, Klaas. (2023) ``Typical = Random'' \textit{Axioms} 12(8): 727. \\ https://doi.org/10.3390/axioms12080727

\vspace{.15cm}
\noindent
Landsman, Klaas. (2022) ``Bohmian Mechanics is Not Deterministic,'' \emph{Foundations of Physics}, \textbf{52}, 73. \\ \href{https://doi.org/10.1007/s10701-022-00611-8}{https://doi.org/10.1007/s10701-022-00611-8}

\vspace{.15cm}
\noindent
Lazarovici, Dustin. (2023) \emph{Typicality Reasoning in Probability, Physics, and Metaphysics}, Cham: Palgrave Macmillan.

\vspace{.15cm}
\noindent
Lazarovici, Dustin and Paula Reichert. (2015) ``Typicality, Irreversibility and the Status of Macroscopic Laws,'' \emph{Erkenntnis}, \textbf{80}, pp.~689--716.

\vspace{.15cm}
\noindent
Lewis, David. (1980) ``A Subjectivist’s Guide to Objective Chance,'' in Richard C. Jeffrey (ed.), \emph{Studies in Inductive Logic and Probability}, Vol. II, Berkeley: University of California Press, 263–293.

\vspace{.15cm}
\noindent
Li, Ming and Paul Vit\'anyi (2019). \emph{An Introduction to Kolmogorov Complexity and Its Applications, 4th Edition}, New York, NY: Springer. 

\vspace{.15cm}
\noindent
Martin-L\"of, Per (1966) ``The definition of random sequences,'' \emph{Information and Control} 9(6):602--619.

\vspace{.15cm}
\noindent
Mierzewski, Krzysztof and Zaffora Blando, Francesca  (2024) ``Two Great Ideas about Chance,'' symposium presentation at the 2024 Philosophy of Science Association Biennial Meeting

\vspace{.15cm}
\noindent
Wilhelm, Isaac. (2022) ``Typical: A Theory of Typicality and Typicality Explanation,'' \emph{The British Journal for the Philosophy of Science}, \textbf{73}(2), pp.~561--581.

\vspace{.15cm}
\noindent
Zaffora Blando, Francesca. (2025) ``Bayesian Merging of Opinions and Algorithmic Randomness,'' \emph{The British Journal for the Philosophy of Science}, 76(4), 921-52. 

\end{document}